 \documentclass[10pt,journal,twocolumn,nofonttune]{IEEEtran}

\usepackage{hyperref}
\usepackage{amssymb,amsmath,latexsym}
\usepackage{graphicx,subfigure}
\usepackage{color,cite,times}
\usepackage{epstopdf}
\usepackage{multirow}
\usepackage{threeparttable}
\usepackage{algorithmic}
\usepackage{array}

\usepackage{amsmath}
\usepackage{bm}
\usepackage[T1]{fontenc}

\usepackage{booktabs}
\usepackage{bbm} 
\usepackage{subfigure}
\usepackage{diagbox}
\def\red#1{{\color{black}#1}}

\newcommand{\nc}{\newcommand}
\nc{\bsm}{\boldsymbol}
\nc{\mbb}{\mathbb}
\nc{\mbs}{\mathbbmss}
\nc{\mbf}{\mathbf}
\nc{\mcl}{\mathcal}

\begin{document}

    \author{Xiaojun~Yuan, Ying-Jun~Angela~Zhang,
            Yuanming~Shi, Wenjing Yan, and Hang~Liu

	\thanks{	
	X. Yuan and W. Yan are with the University of Electronic Science and Technology of China. Y.-J. Zhang and H. Liu are with The Chinese University of Hong Kong. Y. Shi is with ShanghaiTech University.
}}

\title{Reconfigurable-Intelligent-Surface Empowered Wireless Communications: Challenges and Opportunities}

\maketitle

\begin{abstract}
Reconfigurable intelligent surfaces (RISs) are regarded as a promising emerging hardware technology to improve the spectrum and energy efficiency of wireless networks by artificially reconfiguring the propagation environment of electromagnetic waves. Due to the unique advantages in enhancing wireless channel capacity, RISs have recently become a hot research topic. In this article, we focus on three fundamental physical-layer challenges for the incorporation of RISs into wireless networks, namely, channel state information acquisition, passive information transfer, and \red{low-complexity robust system design}. We summarize the state-of-the-art solutions and explore potential research directions. Furthermore, we discuss other promising research directions of RISs, including edge intelligence and physical-layer security.

\end{abstract}

\section{Introduction}

Wireless connectivity is a key enabler to support the evolution from ``connected things" to ``connected intelligence" in the future information society. In particular,  ultra-high data rates, ultra-high reliability, ultra-low latency, and extremely massive connectivity are essential for data sensing, collection, transmission, and processing across future wireless networks. This is achieved by the key enabling wireless technologies, including massive multiple-input multiple-output (MIMO), millimeter wave (mmWave), ultra dense networks and AI-empowered wireless networks \cite{letaief2019roadmap}, complemented by new network functionalities such as edge computing, caching, learning and network slicing. The development of existing wireless systems is based on a basic principle that the radio environments are fixed exogenously and cannot be controlled. Combating the detrimental effects of the radio environments has been the main aim of the design of algorithms, protocols, and systems. Moving forward to unleash the full potential of future wireless networks, it is time now to investigate the feasibility of building a controllable and programable radio environment, so that the radio environment
itself becomes a degree of freedom for system optimization.

Reconfigurable intelligent surfaces (RISs), a.k.a., large intelligent metasurfaces (LIMs) \cite{he2019cascaded} and intelligent reflecting surfaces (IRSs) \cite{wu2019intelligent}, have been envisioned to reduce the energy consumption and improve the spectral efficiency of wireless networks by artificially reconfiguring the propagation environment of electromagnetic waves. Specifically, RISs are typically constructed by the planar (or even conformal) artificial metasurfaces consisting of many reflection amplitude/phase shifts, which are adjustable by a smart controller. The RIS-empowered smart radio environments are thus able to control the phases and/or amplitudes of the incident signals, thereby enhancing the desired signals and mitigating the interference signals. As such, RISs have a huge potential to revolutionize the design of wireless networks, particularly when combined and integrated together with other promising wireless technologies such as ultra-massive MIMO, terahertz communications, AI-empowered wireless networks, and edge intelligence. As such, the purpose of this article is to draw attention to and spur activities on this new research direction.

RIS-empowered smart radio is able to combat the unfavourable propagation conditions (such as deep fading) by manipulating the radio propagation environment in future-generation wireless communications. \red{Millimetre wave and Terahertz communications have been envisioned as enabling technologies for 5G-and-beyond wireless communications.} As the increase of the radio frequency, tens and even hundreds of antenna elements will be installed on base stations and even portable devices. As such, massive MIMO with antenna arrays deployed at both base stations and device terminals is able to provide unprecedented capacity gains to meet the exponential increase in the demand of wireless data services. \red{Meanwhile, however, the increase of the radio frequency weakens the diffraction and scattering effect, rendering electromagnetic waves prone to blockage by obstacles such as buildings in urban areas. As a result, it is difficult to ensure a universal coverage of wireless services in 5G-and-beyond wireless communications using conventional cellular techniques. The recent advancement of the RIS technique provides a revolutionarily new solution to tackle the problem by artificially controlling the propagation environment of electromagnetic waves. Typically, a RIS is composed of a large number of low-cost and energy-efficient reconfigurable reflecting elements that can reflect impinging electromagnetic waves with a controllable phase shift via the help of a smart controller. Through intelligent placement and reflect/passive beamforming, a RIS is able to provide an extra high-quality channel link to overcome the unfavourable propagation conditions of wireless communication systems.}

There are undeniable performance advantages of RIS-empowered smart radio. \red{First, RISs can be deployed almost everywhere to make use of electromagnetic waves that are otherwise dissipated in space. Reconfigurable electromagnetic materials can be used to coat objects in the environment, including but not limited to building facades, ceilings, furnitures, and clothes, etc. Second, RISs are environmentally friendly to meet the requirement of green communications. As the RISs are nearly passive, no additional energy is consumed by RIS-aided systems compared to conventional wireless systems. Third, RISs support full-duplex and full-band transmission, since they only reflect electromagnetic waves. In addition, RISs are cost-effective since they need neither analog-to-digital/digital-to-analog converters nor power amplifiers.} Fourth, the power gain of a RIS follows the quadratic scaling law, as in contrast to the linear power scaling law of a conventional active antenna array. At the same time, however, the use of RISs poses a number of new challenges for the transceiver design of wireless communication systems. In particular, compared with the existing magazine articles \cite{wu2019intelligent,basar2019wireless,di2019smart}, we mainly focus on the fundamental physical-layer problems such as channel state information estimation, passive information transfer, and \red{low-complexity robust system design}. We will also highlight the unique research directions, including RIS-aided edge intelligence and RIS-aided physical-layer security.

The remainder of the article is organized as follows. Section \ref{Sec.channel} discusses channel acquisition in RIS systems. Section III discusses passive information transfer in RIS systems. Robust and low-complexity system design is discussed in Section IV. Section V highlights some of the other research challenges. The article is concluded in Section VI.

\section{Channel State Information Acquisition in RIS-Aided Communication Systems}
\label{Sec.channel}
\subsection{Problem Description}
\label{Sec.channel_1}
\begin{figure}[ht]
  \centering
  \includegraphics[width=3 in]{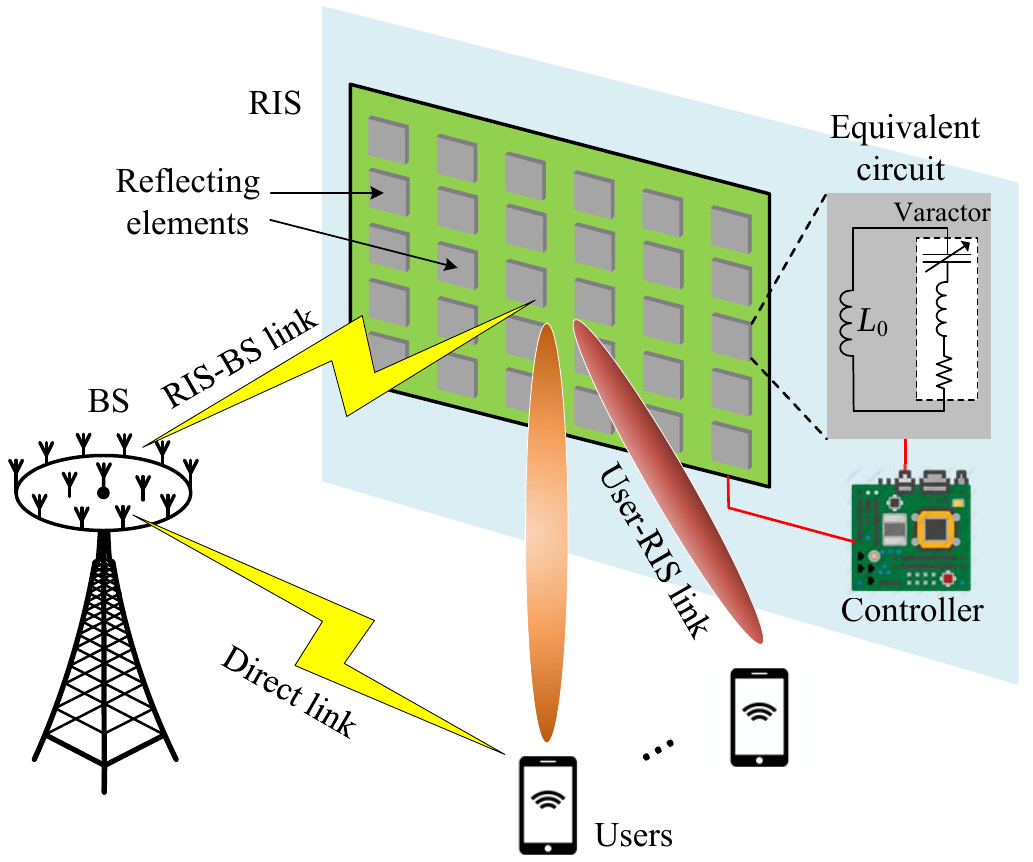}
  \caption{A RIS-assisted massive MIMO system.}\label{Fig.RIS_channel}
\end{figure}

The acquisition of channel state information (CSI) is a fundamental problem to achieve the full potential of RIS-aided wireless networks. Recent studies show that the transceiver design of a RIS-aided system critically depends on the knowledge of CSI, e.g., the joint active and passive beamforming design in \cite{wu2019intelligent} and the joint transmit power allocation and beamforming design in \cite{huang2019reconfigurable}.

The CSI acquisition problem in a RIS-aided system, however, is quite different from that in traditional MIMO communication systems. To be specific, we consider a typical RIS-aided massive MIMO communication system, where a number of users communicate with a multi-antenna base station (BS) via the help of a RIS, as illustrated in Fig.~\ref{Fig.RIS_channel}. In such a system, besides the estimation of the direct channel link (i.e., the channel link between the BS and the users), two additional channel links, namely, the user-RIS channel link and the RIS-BS channel link, also need to be estimated. By turning off all the RIS elements, the direct channel can be obtained based on traditional channel estimation methods. Yet, the remaining problem of estimating the user-RIS and RIS-BS links is far more difficult since the RIS is expected to be a nearly passive device with very limited capability of transmitting, receiving, and processing radio frequency (RF) signals. This means that, unlike conventional pilot-assisted channel estimation, one can neither rely on the RIS to estimate the user-RIS and RIS-BS channel links by processing the pilots from the users and the BS, nor rely on the RIS to transmit pilot signals to facilitate the channel estimation at the users and the BS. As such, CSI acquisition in RIS-aided systems gives rise to a cascaded channel estimation problem, i.e., the problem of estimating the user-RIS and RIS-BS channel links upon the observation of their noisy product. We henceforth refer to the cascade of the user-RIS and the RIS-BS links as the cascaded channel. The cascaded channel estimation problem is in general a bilinear estimation problem, as in contrast to the linear estimation problem for channel estimation in conventional communication systems. In addition, the size of a RIS is usually very large \red{(e.g., in the scale of tens or hundreds)}. This implies that a large number of channel coefficients need to be estimated, which imposes extra difficulty on the cascaded channel estimation problem.

\subsection{State-of-the-Art Solutions}
\label{Sec.CSI_Active}
At present, the design of CSI acquisition in RIS-aided systems is still in its infancy.
Initial attempts to solve this problem can be roughly divided into three categories.

\subsubsection{Active-channel-sensor based CSI acquisition}

This approach is based on the insertion of active channel sensors into the array of passive elements for sensing channel information \cite{taha2019enabling}.
Each active channel sensor is equipped with not only  an RF phase shifter like a passive reflecting element for reflecting the incident electromagnetic (EM) wave, but also an additional baseband processing unit for the channel estimation. Correspondingly, the active sensors have two work modes, namely, the channel sensing mode (using the baseband unit for channel estimation) and the reflection mode (using the RF phase shifters for reflecting EM wave) \cite{taha2019enabling}. During the channel sensing mode, the sensors receive the pilot signals from the users and the BS to estimate their corresponding channel links.
Since the channel coefficients of a large antenna array at the RIS have strong correlation, these coefficients can be constructed based on the sampled channel information by utilizing compressive sensing tools. The channel links from the RIS to the users and to the BS can be obtained by assuming channel reciprocity.

The active-channel-sensor based approach, however, has some disadvantages. First, the active sensors require additional baseband processing units, which increases the hardware cost of the RIS.  Second, the active sensors consume extra energy, which may pose a heavy burden on the RIS as a nearly passive device.
Last but not least, the channel information obtained at the RIS needs to be uploaded to a control centre (usually located at the BS) for beamforming design and resource allocation, which gives rise to the need of information transfer for RIS.
To address the above issues, we next describe two CSI acquisition approaches based on a RIS with all passive elements.

\subsubsection{Channel-decomposition based CSI acquisition}

As discussed in Section~\ref{Sec.channel_1},  the cascaded channel estimation problem arising from RIS-aided systems is difficult to solve due to the multiplication of the two coefficient matrixes of the user-RIS and RIS-BS links.
An idea to tackle this difficulty is to decompose the cascaded channel into a series of sub-channels that are easier to estimate.
For example, one may decompose the cascaded channel into a series of rank-1 matrixes with each corresponding to a RIS element.
Each sub-channel can be estimated by turning on only one RIS element (and turning off all the other elements).  Performing this procedure for each RIS element, the CSI of the whole cascaded channel can be obtained.
This method has been adopted in \cite{mishra2019channel} to estimate the channels in a RIS-aided multiple-input single-output system.
A total pilot length of $N$ is required for reliable CSI acquisition, where $M$ and $N$ are respectively the numbers of the antennas/elements at the BS and at the RIS, respectively. This method can be extended to the multiuser case. However, the required pilot length increases to $NK$ (with $K$ being the number of users), incurring a prohibitively high training overhead when $K$ is large.

A decomposition method for cascaded channel estimation is to estimate the channel by activating each user one by one, i.e., the cascaded channel is decomposed into a series of single-input multiple-output channels seen by each user.
By exploiting the fact that all the users share a common RIS-BS link, the authors in \cite{wang2019channel} proposed an efficient three-phase channel estimation method, which reduces the required pilot length to $K + N + \max\left(K-1, \frac{(K-1)N}{M}\right) $.

\subsubsection{Structure-learning based CSI acquisition}

The cascaded channel of a RIS-aided system usually exhibits strong structural features, such as sparsity and low-rankness, which can be exploited to reduce the overhead of CSI acquisition. The deployment of large-scale antenna arrays enables the RIS and the BS to distinguish EM waves from different directions with high resolution, thereby yielding a sparse representation of the channel matrices in the angular domain (i.e., a large portion of angular channel coefficients are zeros). Likewise, the low-rankness of the cascaded channel matrices arises from the effect of a limited number of scattering paths in the user-RIS and RIS-BS propagation environment. Moreover, signal sparsity can be artificially introduced to assist the cascaded channel estimation by controlling the on/off states of the RIS elements.
With these structural information, the estimation of a cascaded channel can be done by utilizing advanced signal processing tools, such as compressed sensing, sparse matrix factorization,  and low-rank matrix recovery algorithms. For example, the authors in \cite{he2019cascaded} proposed a two-stage algorithm to estimate the cascaded channel in the RIS-aided MIMO system. The algorithm includes a sparse matrix factorization stage to estimate the information of the RIS-BS channel link, and a matrix completion stage to estimate the information of the user-RIS channel link.
\red{Besides channel sparsity, the authors in \cite{liu2020matrix} also exploited the quasi-static property of the RIS-BS channel link
(i.e., the channel components of RIS-BS channel varies much slower than those of the user-BS channel link and the user-RIS channel link since the RIS is usually deployed at a fixed location)
to further reduce the training overhead, and proposed a matrix-calibration-based cascaded channel
estimation method to solve the corresponding estimation problem.
 \red{Table \ref{table1} gives a heuristic comparison of the minimum training lengths of the channel estimation algorithms in \cite{he2019cascaded,wang2019channel} and \cite{liu2020matrix} versus the number of RIS elements.}}

\begin{table}[t]
\red{
\caption{Minimum training lengths of respective channel estimation algorithms versus the number of RIS elements.}
\centering
\begin{threeparttable}
\vspace{-5mm}
\begin{tabular}{|l|c|c|c|c|c|c|c|}
\hline\hline
\multirow{2}{*}{\diagbox{Algorithm}{\!\!\!\!\!\!\!\!\!\!\!\!\!\! No. of RIS elements\!\!} } & \multirow{2}{*}{\!$128$\!} & \multirow{2}{*}{\!$200$\!\!} & \multirow{2}{*}{\!$256$} & \multirow{2}{*}{\!\!$300$\!\!} & \multirow{2}{*}{\!$400$\!} \\
& & & & & \\
\hline
Method of \cite{mishra2019channel}  & \!\!\!$12800$\!\!\! & \!\!\!$20000$\!\!\! & \!\!\!$25600$\!\!\! & \!\!\!$30000$\!\!\! & \!\!\!$40000$\!\!\! \\
\hline
Method of \cite{he2019cascaded}  & \!$245$\! & \!$324$\! & \!$438$\! & \!$468$\! & \!\!$981$\!\! \\
\hline
Method of \cite{wang2019channel} & \!$227$\! & \!$299$\! & \!$383$\! & \!$449$\! & \!$598$\!\\
\hline
Method of \cite{liu2020matrix}   &\!$53$\! &\!$55$\! & \!$58$\!& \!$65$\!&\!$83$\!  \\
\hline
Analytical lower bound in \cite{liu2020matrix}  &\!$20$\!& \!$36$\! & \!$40$\!& \!$47$\!&\!$62$\! \\
\hline\hline
\end{tabular}
\vspace{-2pt}
\begin{tablenotes}
        \item[a] We consider a RIS-aided uplink MIMO system where the number of antennas at the BS and the number of single antenna user are $200$ and $100$, respectively. The noise variance is set to $0$. For the method of \cite{he2019cascaded}, the rank of the user-RIS channel is $5$. The minimum training lengths of respective algorithms are obtained when the channel MSEs are less than $-60\,\text{dB}$. The active rate of the RIS elements required in \cite{he2019cascaded} is set to be $0.15$.
\end{tablenotes}
\end{threeparttable}
\label{table1}  }
\end{table}

\subsection{Research Challenges}
\subsubsection{Channel modelling and channel acquisition}

Channel modelling of RIS-aided MIMO systems has not yet been well understood. A conventional MIMO channel is usually assumed to be far-field, where EM waves impinge upon an antenna array nearly in parallel under the assumption that the radio source, the scatterers, and the receiver are located sufficiently apart from each other. However, the passive antenna array of a RIS, coated on a facade of a building or on the ceiling of a room, can have a size that is comparable with its distance from the base stations or the mobile devices. As such, it is necessary to take into account
near-field propagation properties in modelling the BS-RIS-user channels. Other propagation properties, such as line-of-sight (LOS)/non-LOS/narrow-band/broadband, etc., may also be radically different for RIS-aided MIMO systems as compared to conventional MIMO systems. Therefore, new models are needed to characterise the propagation environment of a RIS-aided MIMO system more precisely.

The new models also encourage the use of new mathematical tools in channel acquisition. For example, matrix factorization and matrix completion techniques are involved in the cascaded channel estimation algorithm developed in \cite{he2019cascaded}. Yet, the channel model employed in  \cite{he2019cascaded} is very primitive. We believe that with more realistic channel modelling, other advanced statistical signal processing techniques, such as tensor factorization and structured signal reconstruction, will find their roles in reliable acquisition of the CSI.

\subsubsection{System design under CSI uncertainty}
\label{Sec.CSI_uncertainty}

The existing studies on the design of RIS-aided systems are mostly based on the assumption of perfect CSI, so that the phases of the reflecting elements of the RIS can be judiciously adjusted for performance enhancement. As mentioned previously, in practice, CSI acquisition of the cascaded channel links is a difficult problem due to limited signal processing capabilities of the RIS. As such,
the design of RIS-aided systems including transceiver design and passive beamforming optimization at the RIS needs to be carried out under CSI uncertainty. \red{It is worth mentioning that the outdated CSI problem, as a model of CSI uncertainty due to CSI acquisition delay, is more likely to occur in RIS-aided systems. The reason is that, besides the CSI of the direct link, a large amount of the channel coefficients of the user-RIS and RIS-BS links need to be delivered between the transceiver and the RIS, which causes additional delay. This inspires us to investigate the design of RIS-aided systems under outdated CSI.}

In addition, it is known that in wireless communication systems, joint channel estimation and signal detection provides significant performance improvement over separate processing approaches. Thus, it is desirable to investigate the possibility of jointly estimating the user-RIS and RIS-BS channels and at the same time detecting the data from the users. This is a highly non-linear signal estimation problem that calls for urgent solutions.

\subsubsection{Theoretical limits}

The fundamental performance limit of RIS-aided communication systems is far from being well understood. For example, for a general RIS-aided massive MIMO system, it is so far not clear how many pilots are required to reliably estimate the three channel links; it is also unknown that by exploiting the additional channel structures (such as sparsity and low-rankness), how much pilot reduction can be achieved. An explicit characterization of the fundamental tradeoff between the training overhead and the system parameters (such as $M$, $N$, and $K$) is highly desirable.


\section{Passive Information Transfer of RIS}

\subsection{Why Passive Information Transfer?}

The existing studies mostly focused on the utilization of RISs to enhance the primary end-to-end communications by performing passive beamforming.
However, in practice, RISs also need to transfer information. The potential sources of the RIS information are listed as follows.

\begin{itemize}

\item {\it Control signaling of RIS:} To coordinate with the transceiver, a RIS is required to report its state information in real time. For example, to synchronize with the transceiver for packet delivery, the RIS needs to acknowledge its current status by sending out control signals.

\item {\it Maintenance of RIS:}  It is important to monitor the environmental conditions (such as temperature, humidity, pressure, etc.) of the RIS in real time to guarantee its normal operation. In addition, if some elements of the RIS are impaired, such information needs to be reported to the control center.

\item {\it Assistance of CSI acquisition:} As mentioned in Section \ref{Sec.CSI_Active}, with active-channel-sensor based CSI acquisition, the channel is estimated at the RIS based on the received signals of the inserted active sensors.
 The CSI acquired at the RIS  needs to be forwarded to the transmitter for beamforming design.

\item {\it Green IoT:} Hardware cost and energy consumption are fundamental bottlenecks to constrain the extensive implementation of IoT devices. The combination of IoT with backscatter communication is regarded as a promising solution to overcome these obstacles and achieve green IoT \cite{liu2013ambient}.
    Compared with backscattering, RISs are able to reflect incident signals in a much more efficient way. This inspires the use of RISs to assist information transfer of IoT devices.

\end{itemize}

Based on the above discussions, information transfer at RISs is a critical problem for the development of the RIS technology. An immediate solution to this problem is to equip each RIS with a dedicated transmitter. However, this solution is not cost-effective and requires extra power consumption. Instead, it is more desirable to modulate the RIS information onto its reflected signals to achieve passive information transfer.
We next briefly introduce the state-of-the-art research along this line.

\subsection{State-of-the-Art Solutions}

In \cite{basar2019transmission}, the RIS is regarded as an access point by assuming that the RIS is supported by a nearby radio
frequency (RF) signal generator. To transmit RIS information, the RF signal generator emits an unmodulated carrier signal to the RIS, and the RIS
modulates its information onto the reflected carrier signal. Simultaneously, the RIS is required to maximize the received signal-to-noise ratio (SNR) by manipulating the phase shifts of the RIS. However, the use of the dedicated RF signal generator is not cost-effective.

In \cite{yan2019passive}, the authors proposed a joint passive beamforming and information transfer (PBIT) technique for the RIS-aided communication systems, which aims to simultaneously transmit the RIS information and enhance the primary communication quality. Furthermore,
the authors proposed to adopt the spatial modulation on the index of the RIS elements to transmit the RIS information in a completely passive manner. That is, the RIS information is transmitted by manipulating the on/off states of the RIS elements.
 Compared to \cite{basar2019transmission}, the PBIT scheme in \cite{yan2019passive} is more promising since it does not consume any extra time/frequency resource in information transfer.

\subsection{Research Challenges}
\begin{figure}[t]
  \centering
  \includegraphics[width=3 in]{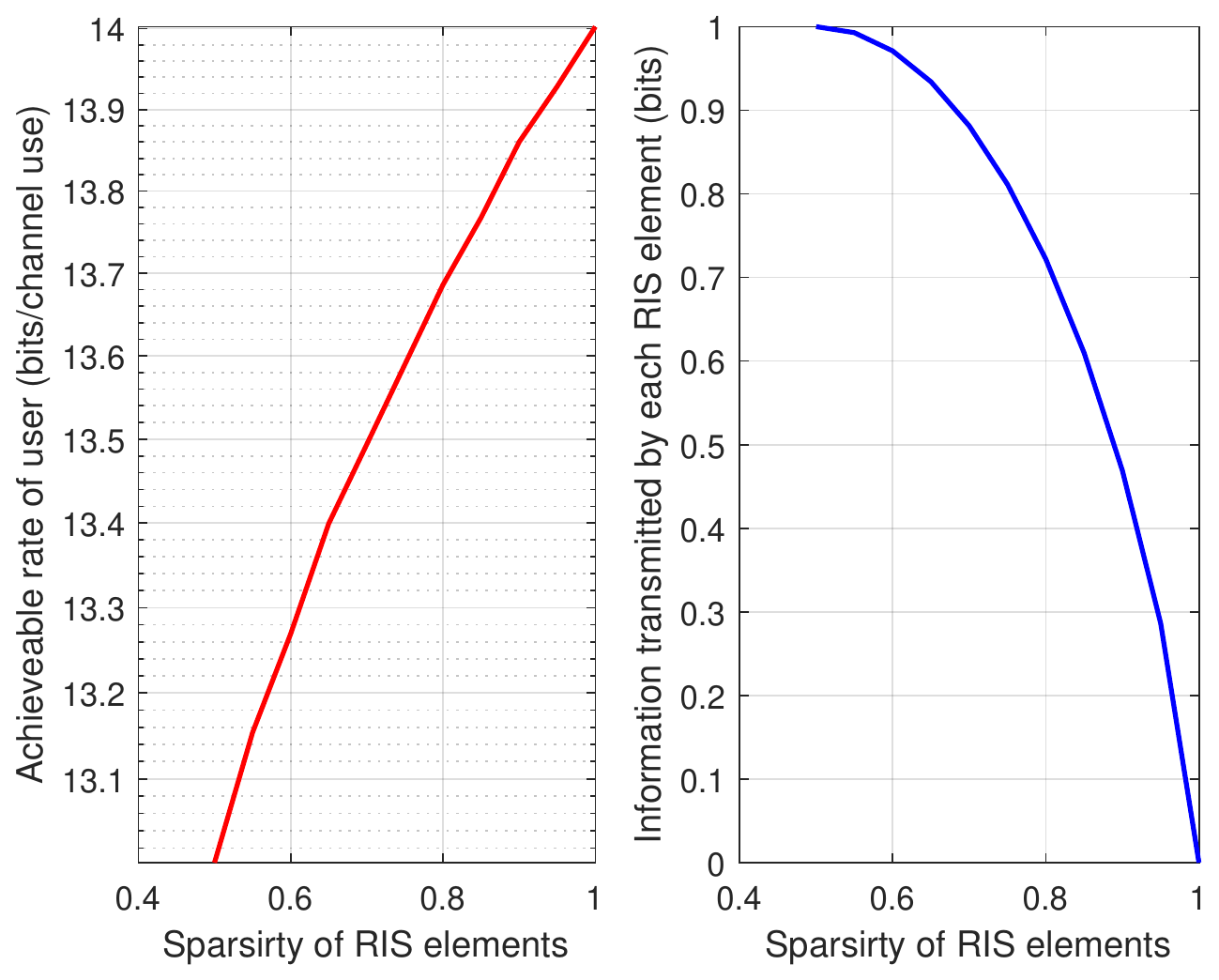}
  \red{\caption{An illustration of the tradeoff between the achievable rate of the primary system and the amount of information carried by each RIS element in a RIS-aided SIMO system, where the numbers of the receive antennas and the RIS elements are both set to 128. The phase shifts of the RIS elements are randomly taken as a complex number with unit modulus. All the channel coefficients are independently drawn from the standard circularly symmetric complex Gaussian distribution.} \label{Fig_PBIT}}
\end{figure}

Passive information transfer of RIS, especially its joint design with passive beamforming, is an emerging research direction rich of open challenges as discussed below.

\subsubsection{RIS design}

In the PBIT scheme, the RIS is required to enhance the primary communication and simultaneously deliver its private information.
Then, how to balance these two functionalities becomes an essential problem for the design of RIS.
 A straightforward approach is to divide all the RIS elements into two groups, one for performing passive beamforming and the other for transferring information. One disadvantage of this approach is that the RIS elements used to transfer information do not contribute to enhancing the primary communication. To address this issue, the authors in \cite{yan2019passive} proposed to enable simultaneous passive beamforming and information transfer at each RIS element, where spatial modulation is applied to each RIS element for information delivery. The spatial modulation method randomly turns off a portion of RIS elements for information delivery, which generally compromises the capabilities of enhancing the primary communication at the RIS. \red{A simple illustration of the tradeoff between the achievable rate of the primary system and the amount of information carried by each RIS element in a RIS-aided single input multiple output (SIMO) system under the PBIT scheme \cite{yan2019passive} is given in Fig. \ref{Fig_PBIT}. We see that, with the sparsity of the RIS elements (i.e., the probability of each RIS element being turned on) increases from $0.5$ to $1$, the achievable rate of the primary system increases from about $13$ bits to $14$ bits, while the information carried by each RIS element deceases from $1$ bit to $0$. A deeper understanding of the tradeoff between the passive information transfer capability and the passive beamforming gain calls for urgent investigation.}


 The passive beamforming design in the PBIT scheme generally involves stochastic optimization since the carried information introduces randomness on the RIS reflecting coefficients. Solving stochastic optimization problems is far more difficult than solving deterministic optimization problems involved in traditional beamforming design. In addition, various design criteria in terms of spectrum and power efficiency shall be considered in the problem formulations.

%

\subsubsection{Joint transceiver and RIS design}

We start with the transmitter side. In a RIS-assisted massive MIMO system, the active beamforming design at the transmitter needs to be optimized together with the passive beamforming at the RIS to achieve globally optimal system performance. In the PBIT scheme, joint active and passive beamforming design is particularly challenging due to the need of stochastic optimization caused by the randomness of RIS information.

The receiver of the PBIT scheme is required to retrieve the information from both the RIS and the transmitters. The signals from the transmitters and the RISs are multiplied together, resulting in bilinear signal detection problems. This inspires the development of new statistical inference techniques for bilinear models.

From an information theoretic perspective, the PBIT system can be modelled by a multiplicative multiple access channel. The capacity of such a channel is not well understood so far. The joint design of the channel coding and beamforming strategies at the transmitter and the RIS, together with the detection and decoding algorithms at the receiver poses a highly challenging task worthy of further investigation.

\section{\red{Low-Complexity Robust System Design}\label{sec:resource}}

\red{The unprecedented capacity gain brought by RISs comes at a high computational cost. As discussed in Section \ref{Sec.CSI_Active}, to significantly reduce the training overhead, Bayesian inference methods are needed to directly factorize the cascaded channel matrices, yielding much higher computational complexity than the traditional channel estimation methods (e.g., in \cite{mishra2019channel}).
Likewise, the joint optimization of RIS phase shifts and transceiver design leads to large-scale non-convex optimization problems, which are difficult to solve especially when the size of the RIS grows. This section discusses the tradeoff between the system performance and the computational cost. In particular, we conjecture that due to the large number of RIS elements, the system performance is often robust against the imperfect phase tuning of individual elements.}

\begin{figure}[t]
  \centering
  \includegraphics[width=0.4\textwidth]{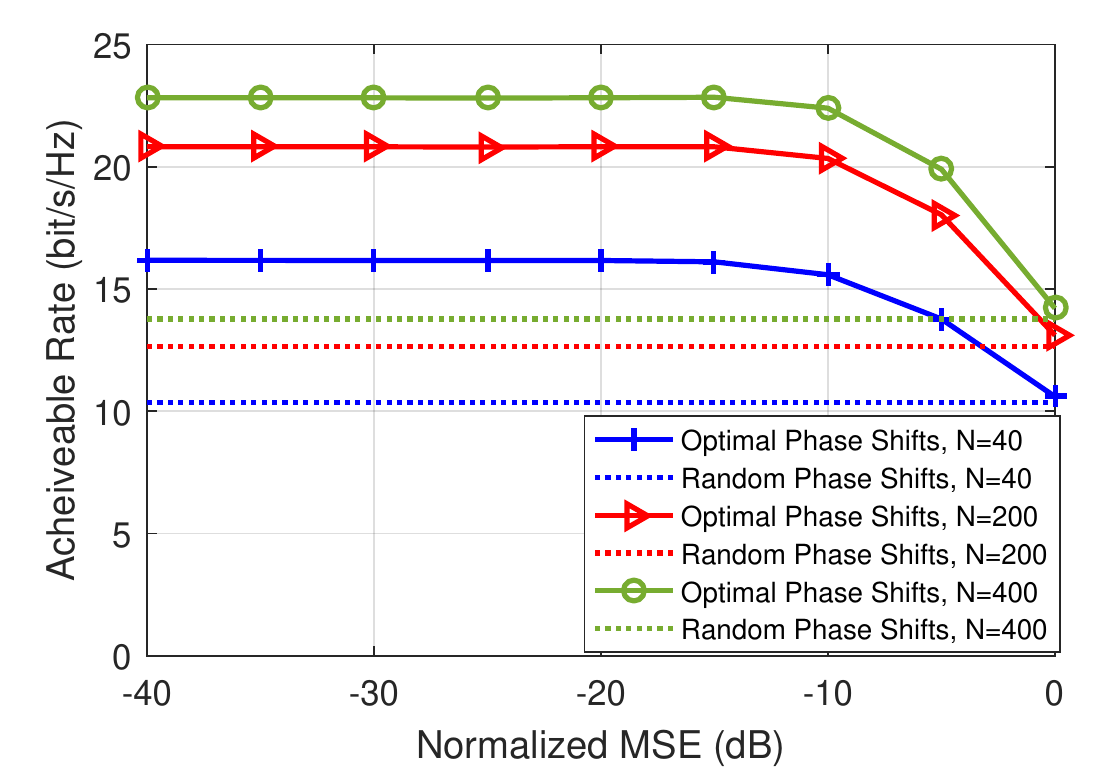}\\
  \red{\caption{Achievable rate versus channel estimation error in a RIS assisted system. A BS with $M=40$ antennas communicates with a single-antenna user with the help of a RIS, where the number of RIS elements $N$ ranges from $40$ to $400$. Solid lines represent the cases of optimal phase shifts, where the RIS phase shifts are optimized with respect to estimated channel information. Dotted lines represent the cases of random phase shifts, where the RIS phase shifts are randomly selected in $[0, 2\pi]^N$.}
  \label{fig:robust}}
\end{figure}

\begin{figure}[t]
  \centering
  \includegraphics[width=0.4\textwidth]{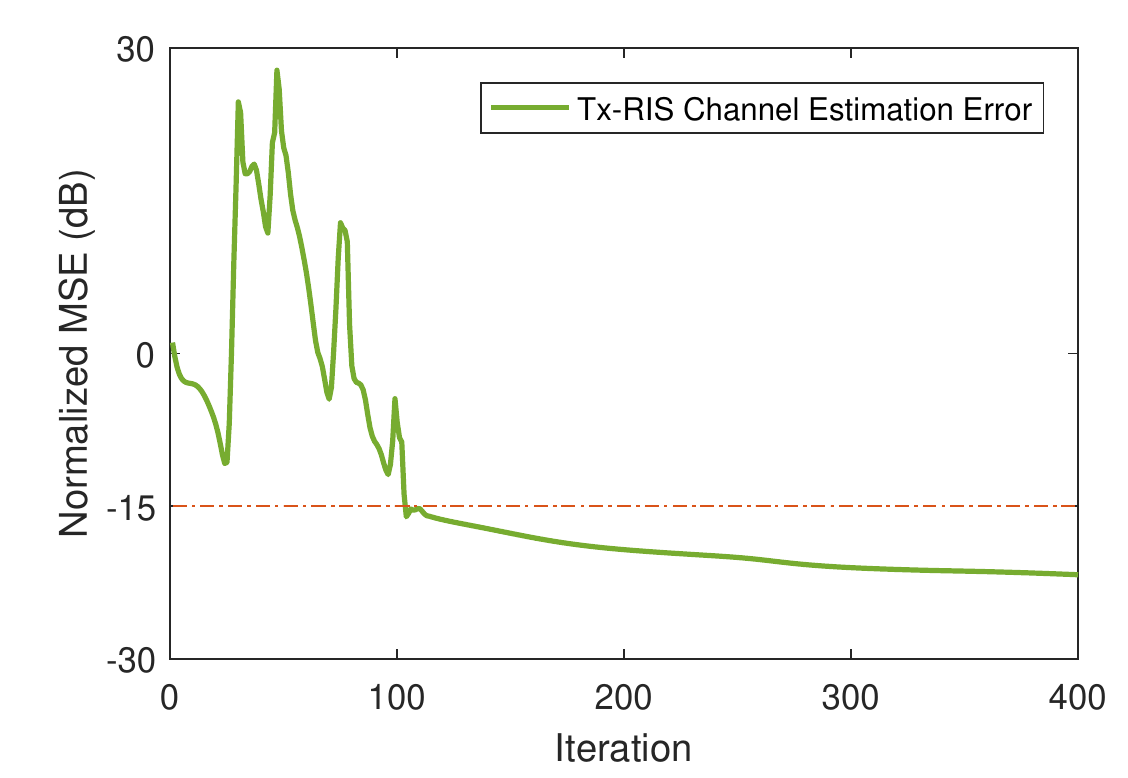}
  \red{\caption{MSE vs. message passing iterations in the matrix calibration based cascaded channel estimation algorithm \cite{liu2020matrix}.}
  \label{fig:iteration}}
\end{figure}

\red{\subsection{Robustness Against Channel Estimation Errors}}

\red{Fig. \ref{fig:robust} illustrates the robustness of the achievable data rate against the channel estimation error. This figure shows that the loss of data rate is negligible when the normalized mean square error (MSE) is as large as $-10$ dB. Intuitively, this is because a large number of RIS elements offsets the need of accurate phase calibration. The relatively large error tolerance would allow us to put the channel estimation algorithm to an early stop when the MSE drops to an acceptable level. To see this, Fig. \ref{fig:iteration} depicts the convergence of the message passing algorithm for Bayesian matrix factorization in \cite{liu2020matrix}. The figure shows that the algorithm achieves $-15$dB normalized MSE very quickly within the first 100 iterations. However, to further increase the accuracy to $-20$dB, more than 400 iterations are needed. Considering the performance robustness in Fig. \ref{fig:robust}, we can safely stop the algorithm at the $100$-th iteration without waiting for the algorithm to fully converge.}

\red{The above observation naturally leads to the question: \emph{How much computational cost shall we spend on cascaded channel estimation?} To answer this question, it is critical to seek the following fundamental understandings: 1) How to characterize the relationship between the performance metrics (such as achievable data rate and outage probability) and CSI accuracy. 2) How to analytically track the MSE evolution in the iterative Bayesian inference algorithms. For example, \cite{liu2020matrix} borrowed the replica method from statistical physics to analyze the MSE evolution in the  matrix factorization algorithm therein.}

\red{\subsection{Robustness Against Low-Resolution Phase Shifts}}

\red{In practice, the RIS phase shifts can only take discrete values due to the finite quantization levels of the hardware, rendering the RIS optimization problem an intractable mixed-integer non-convex optimization problem. To reduce the hardward cost and control signaling overhead, recent studies have investigated the robustness of system performance with respect to low-resolution quantization. It has been shown that the capacity degradation is below 1bit/s/Hz when the number of quantization bits is reduced from infinity to 2.  Indeed, analytically characterizing the impact of low-resolution quantization on the overall system performance will be a challenging yet important research topic.}

\red{Low-resolution quantization allows the design of low-complexity optimization algorithms due to the limited number of candidate solutions. In addition to conventional integer programming algorithms, low-resolution quantization opens the possibility of designing reinforcement-learning based algorithms with the action space greatly reduced. For example, \cite{huang2019deep} recently proposed an actor-critic based mixed integer programming solver, where the actor network adopts a deep neural network to learn the integer variables and the critic network uses mathematical optimization to tackle the continuous variables.  The solver is shown to be very effective when the action space is small.}

\section{Other Challenges of RIS-Aided Wireless Communications}

\subsection{RIS-Aided Edge Intelligence}

Edge intelligence, including edge caching, edge computing, and edge learning, is an advanced technology to relieve the data traffic of networks by utilizing the storage units at edge servers, to solve the computation latency problem of highly demanding devices via computation offloading, and to guarantee the privacy and security of big data analysis by adding computing and learning functionalities into radio access networks. However, the utilization of edge intelligence heavily relies on the inhabited network topology and the limited energy budget of edge devices. To design communication-efficient content delivery strategies for edge caching network and data shuffling strategies for edge computing system, the RIS provides a promising solution to improve the achievable degree-of-freedoms via coping with rank-deficient channels and mitigating the co-channel interference. This is achieved by improving the feasibility of the interference alignment conditions via actively controlling the network environments. \red{The RIS can also boost the received signal power for over-the-air computation in the user scheduling procedure, thereby enabling low-latency global model aggregation in edge learning. This is achieved by intelligently tuning the phases of the incident EM waves and exploiting the waveform superposition property of a wireless multiple-access channel to adapt to the local model updates, thereby improving the scheduling policy for fast edge learning.}

%
%

\subsection{RIS-Aided Physical-Layer Security}

The security of wireless networks is of crucial importance since wireless networks have been used increasingly for a wide variety of security-sensitive applications, including but not limited to banking, social networking, and environmental surveillance. Recently, there has been a rising interest in the development of secure data transmission based on physical properties of the wireless channel (hence the name physical-layer security). The use of RISs provides a radically new mechanism in which the propagation environment around the insecure nodes is manipulated to prevent potential information leakage and ensure network security. \red{This is achieved by smartly transforming or recycling the existing signals via programming the wireless channel propagations to enhance the signals for the legitimate users and cancel out the signals to eavesdroppers.} Mathematically, the integration of RISs with physical-layer security brings about new communication models, and hence new optimisation problems. These problems are generally non-convex and difficult to solve. As such, it is a pressing challenge to develop new optimisation techniques for solving these problems. \red{Furthermore, since the delay and interference in the system may yield outdated cascaded channel acquisition, it becomes critical to characterize the performance of the RIS-aided physical layer security under outdated channel state information \cite{le2018secrecy}.}

\section{Conclusions}

This article discussed the fundamental  \red{physical-layer} issues related to the deployment of RISs in practical wireless networks, including CSI acquisition, passive information transfer, and  \red{low-complexity robust system design}. For each of these issues, we explained the main challenges, discussed the state-of-the-art solutions, and pointed out the open research directions. Other RIS design problems, such as edge intelligence and physical-layer security were briefly introduced. This article serves as a humble attempt to provide useful guidance and insightful inspiration to future research endeavours on RIS-aided wireless communication systems. We believe that the the RIS technique, due to its superb advantages in enhancing wireless communications, will play an indispensable role in the 5G-and-beyond era.

\bibliographystyle{IEEEtran}
\bibliography{magazine}

\begin{thebibliography}{10}
\providecommand{\url}[1]{#1}
\csname url@samestyle\endcsname
\providecommand{\newblock}{\relax}
\providecommand{\bibinfo}[2]{#2}
\providecommand{\BIBentrySTDinterwordspacing}{\spaceskip=0pt\relax}
\providecommand{\BIBentryALTinterwordstretchfactor}{4}
\providecommand{\BIBentryALTinterwordspacing}{\spaceskip=\fontdimen2\font plus
\BIBentryALTinterwordstretchfactor\fontdimen3\font minus
  \fontdimen4\font\relax}
\providecommand{\BIBforeignlanguage}[2]{{%
\expandafter\ifx\csname l@#1\endcsname\relax
\typeout{** WARNING: IEEEtran.bst: No hyphenation pattern has been}%
\typeout{** loaded for the language `#1'. Using the pattern for}%
\typeout{** the default language instead.}%
\else
\language=\csname l@#1\endcsname
\fi
#2}}
\providecommand{\BIBdecl}{\relax}
\BIBdecl

\bibitem{letaief2019roadmap}
K.~B. Letaief, W.~Chen, Y.~Shi, J.~Zhang, and Y.-J.~A. Zhang, ``The roadmap to
  6{G}: {AI} empowered wireless networks,'' \emph{IEEE Commun. Mag.}, vol.~57,
  no.~8, pp. 84--90, Aug. 2019.

\bibitem{he2019cascaded}
Z.-Q. He and X.~Yuan, ``Cascaded channel estimation for large intelligent
  metasurface assisted massive {MIMO},'' \emph{IEEE Wireless Commun. Lett.},
  vol.~9, no.~2, pp. 210--214, Feb. 2019.

\bibitem{wu2019intelligent}
Q.~Wu and R.~Zhang, ``Towards smart and reconfigurable environment: Intelligent
  reflecting surface aided wireless network,'' \emph{IEEE Commun. Mag.},
  vol.~58, no.~1, pp. 106--112, Jan. 2020.

\bibitem{basar2019wireless}
W.~Tang \emph{et~al.}, ``Wireless communications with programmable metasurface:
  New paradigms, opportunities, and challenges on transceiver design,''
  \emph{IEEE Wireless Commun.}, vol.~27, no.~2, pp. 180--187, Apr 2020.

\bibitem{di2019smart}
M.~Di~Renzo \emph{et~al.}, ``Smart radio environments empowered by {AI}
  reconfigurable meta-surfaces: {A}n idea whose time has come,'' \emph{EURASIP
  J. Wirel. Commun. Netw.}, vol. 2019, no.~1, pp. 1--20, May 2019.

\bibitem{huang2019reconfigurable}
C.~Huang, A.~Zappone, G.~C. Alexandropoulos, M.~Debbah, and C.~Yuen,
  ``Reconfigurable intelligent surfaces for energy efficiency in wireless
  communication,'' \emph{IEEE Trans. Wireless Commun.}, vol.~18, no.~8, pp.
  4157--4170, Aug. 2019.

\bibitem{taha2019enabling}
\BIBentryALTinterwordspacing
A.~Taha, M.~Alrabeiah, and A.~Alkhateeb, ``Enabling large intelligent surfaces
  with compressive sensing and deep learning,'' Apr. 2019. [Online]. Available:
  \url{https://arxiv.org/abs/1904.10136}
\BIBentrySTDinterwordspacing

\bibitem{mishra2019channel}
D.~Mishra and H.~Johansson, ``Channel estimation and low-complexity beamforming
  design for passive intelligent surface assisted {MISO} wireless energy
  transfer,'' in \emph{Proc. IEEE Int. Conf. Acoust. Speech Signal Process.
  (ICASSP)}.\hskip 1em plus 0.5em minus 0.4em\relax Brighton, UK, 2019, pp.
  4659--4663.

\bibitem{wang2019channel}
Z.~Wang, L.~Liu, and S.~Cui, ``Channel estimation for intelligent reflecting
  surface assisted multiuser communications: Framework, algorithms, and
  analysis,'' \emph{IEEE Trans. Wireless Commun.},
  {DOI:}10.1109/TWC.2020.3004330.

\bibitem{liu2020matrix}
H.~Liu, X.~Yuan, and Y.-J.~A. Zhang, ``Matrix-calibration-based cascaded
  channel estimation for reconfigurable intelligent surface assisted multiuser
  {MIMO},'' \emph{IEEE J. Sel. Areas Commun.}, 10.1109/JSAC.2020.3007057.

\bibitem{liu2013ambient}
V.~Liu \emph{et~al.}, ``Ambient backscatter: wireless communication out of thin
  air,'' in \emph{ACM SIGCOMM}, vol.~43, no.~4.\hskip 1em plus 0.5em minus
  0.4em\relax ACM, Aug. 2013, pp. 39--50.

\bibitem{basar2019transmission}
E.~Basar, ``Transmission through large intelligent surfaces: {A} new frontier
  in wireless communications,'' in \emph{Proc. Eur. Conf. Netw. Commu.
  (EuCNC)}.\hskip 1em plus 0.5em minus 0.4em\relax Valencia, Spain, 2019, pp.
  112--117.

\bibitem{yan2019passive}
W.~Yan, X.~Yuan, and X.~Kuai, ``Passive beamforming and information transfer
  via large intelligent surface,'' \emph{IEEE Wireless Commun. Lett.}, vol.~9,
  no.~4, pp. 533--537, Apr. 2020.

\bibitem{huang2019deep}
L.~Huang, S.~Bi, and Y.~J. Zhang, ``Deep reinforcement learning for online
  computation offloading in wireless powered mobile-edge computing networks,''
  \emph{IEEE Trans. Mobile Comput.}, {DOI:}10.1109/TMC.2019.2928811.

\bibitem{le2018secrecy}
K.~N. Le, ``Secrecy and end-to-end analyses employing opportunistic relays
  under outdated channel state information and dual correlated rayleigh
  fading,'' \emph{IEEE Trans. Veh. Technol.}, vol.~67, no.~11, pp.
  10\,504--10\,518, Nov. 2018.

\end{thebibliography}

\textbf{Xiaojun Yuan} [S'04-M'09-SM'15] (xjyuan@uestc.edu.cn) received his Ph.D. degree in Electrical Engineering from the City University of Hong Kong. He is now a professor with the University of Electronic Science and Technology of China, supported by the Thousand Youth Talents Plan in China.

\textbf{Ying-Jun Angela Zhang} [S'00-M'05-SM'10-F'20] (yjzhang@ie.cuhk.edu.hk) received her Ph.D. degree from The Hong Kong University of Science and Technology. She is now an Associate Professor at the Department of Information Engineering, The Chinese University of Hong Kong.

\textbf{Yuanming Shi} [S'13-M'15] (shiym@shanghaitech.edu.cn) received his B.S. degree from Tsinghua University and the Ph.D. degree from The Hong Kong University of Science and Technology. He is currently a tenured Associate Professor at the School of Information Science and Technology,
ShanghaiTech University.

\textbf{Wenjing Yan} [S'20] (wjyan@std.uestc.edu.cn) received the B.S. degree in Chongqing University, in 2018. She is currently working toward the M.S. degree in the University of Electronic Science and Technology of China.

\textbf{Hang Liu} [S'19] (lh117@ie.cuhk.edu.hk) received the B.Sc. degree in Mathematics and Information Engineering from The Chinese University of Hong Kong, Hong Kong, in 2017, where he is currently pursuing the Ph.D. degree.

\end{document}